\documentclass[12pt,a4paper,notitlepage,oneside,dvips,openany]{article}
\usepackage{epsf,rotate}
\usepackage{amsmath,amsfonts,amstext,amsbsy,amsopn,amsfonts,amssymb}

\newcommand{\mysection}[1]{\section{#1\setcounter{equation}{0}}}
\renewcommand{\theequation}{\arabic{section}.\arabic{equation}}
\newcommand{\eval}[2][\right]{\relax
  \ifx#1\right\relax \left.\fi#2#1\rvert}
\newcommand{\Log}{\mbox{Log}}
\newcommand{\Arg}{\mbox{Arg}}
\newcommand{\RRe}{\mbox{Re}}
\newcommand{\IIm}{\mbox{Im}}
\newcommand{\dz}[1]{\frac{d{#1}}{dz}}

\newcommand{\px}[2]{\frac{\partial {#1}}{\partial {#2}}}

\begin{document}

\title{\hfill TPJU-11/2001\\
~~\\
Solutions of the quantization conditions
for the odderon charge $q_3$ and conformal weight $h$}
\author{{\bf Jan Kota{\'n}ski and Micha\l{}  Prasza\l{}owicz}  \\
{\em M.Smoluchowski Institute of Physics,} \\
{\em Jagellonian University,} \\
{\em Reymonta 4, 30-059 Krak{\'o}w, Poland.}}

\date{November 14, 2001}
\maketitle

\begin{abstract}
The quantization conditions which come from the requirement of the
singlevaluedness of the odderon wave function  are formulated and
solved numerically in the 4 dimensional space of the odderon
charge $q_3$ and the conformal weight $h$. It turns out that these
conditions are fulfilled along one dimensional curves parametrized
by a discrete set of values of $\RRe~h$ in 3 dimensional subspace
$(\IIm~h,\IIm~q_3,\RRe~q_3)$. The odderon energy calculated along
these curves corresponds in all cases to the intercept lower than
1.
\end{abstract}

\mysection{Introduction}

The leading contribution to the total elastic
scattering amplitude of two hadrons (A, B)
can be written in Quantum Chromodynamics (QCD) in a so called Regge limit
\begin{equation}
s\rightarrow \infty, \;\; t=const.
\label{eq:rlim}
\end{equation}
as a power series in a strong coupling constant $\alpha_s$ of the partial
amplitudes with a given number $n$ of  reggeized gluons (Reggeons) propagating
in the $t$ channel:
\begin{equation}
\begin{aligned}
A(s,t)&=\sum_{n=2}^{\infty} \alpha_s^{n-2} A_n(s,t), \\
A_n (s,t)&=i \sum_{\{\alpha \}}
\beta_A^{n,\varkappa_n} (t) \beta_B^{n,\varkappa_n} (t) s^{\alpha_n(t)}.
\label{eq:An}
\end{aligned}
\end{equation}
Here $\varkappa_n$ denotes quantum numbers of the $n$ Reggeon state
and the residue functions $\beta_A^{n,\varkappa_n}$ and $\beta_B^{n,\varkappa_n}$
measure the overlap between the hadronic wave functions and the
wave function of a compound state of $n$ reggeized gluons.
The $n$-Reggeon's partial amplitudes are proportional to $s^{\alpha_n(t)}$ where
$\alpha_n(0)$ is called an intercept.

It is of great importance to calculate the intercepts
$\alpha_n(0)$ in QCD, not only because they govern the high energy
behavior of the forward elastic amplitudes but also because, {\em
e.g.} for $n=2$, they are responsible for the small Bjorken $x$
behavior of the deep inelastic structure functions. The lowest
non-trivial contribution for $n=2$ was calculated in the leading
logarithmic approximation by Balitsky, Fadin, Kureav and Lipatov
\cite{KLF,BL}, who derived and solved  equation for the Pomeron
intercept. The equation for three and more Reggeons was formulated
in Refs.\cite{Bart,kwprasz,jar}. It took, however, almost 20 years
before the solution for $n=3$ was obtained in Refs.\cite{jw2,jw1}.

The real progress started with an observation that the $n$-Reggeon
exchange is equivalent to an eigenvalue problem of a Schr\"odinger
like equation with calculable interaction hamiltonian $\Hat{\cal
H}_n$. Here the eigen-energy is related to $\alpha_n(0)-1$. This
Schr\"odinger problem is exactly solvable \cite{lip,lip3,fadkorch}
which means that there exist $n-1$ integrals of motion
$(\hat{q}_2,\ldots , \hat{q}_n)$ which commute with $\Hat{\cal
H}_n$ and among themselves. The eigenvalue of $\hat{q}_2$ is equal
$-h(h-1)$ where $h$ is called a conformal weight.

In the present work we shall concentrate the odderon exchange,
{\em i.e.} on the case with $n=3$. It is easier to conduct the calculations
in the impact parameter space,
that is in the transverse spatial coordinates of $n$ Reggeons
$(x_j, y_j)$. After introducing the complex coordinates ($z_j=x_j+y_j$,
$z_j^\ast=x_j-y_j$) for $j$-th reggeized gluon,
the odderon Hamiltonian becomes holomorphically separable
\begin{equation}
\Hat{\cal H}_3=\Hat{H}_3+\Hat{\overline{H}}_3=
\frac{\alpha_s N_c}{4 \pi} \sum_{k=1}^{3}
\left[
\Hat{H}(z_k,z_{k+1})+\Hat{H}(z^{\ast}_k,z^{\ast}_{k+1})
\right]
\end{equation}
where $N_c$ is a number colors and $z_1=z_{n+1}$.
The Hamiltonian $\Hat{\cal H}_3$ is conformally invariant. Its eigenfunction
is given as a bilinear form $\varPhi=\overline{\varPsi} \times \varPsi$
where $\varPsi$ ($\overline{\varPsi}$) is the solution of the Schr{\"o}dinger
equation in the holomorphic (antiholomorphic) sector.

There have been many attempts either to directly find the values
of $E_3$ \cite{GLN}\nocite{abr1,abr2,br2}-\cite{korch2} or to find
the spectrum of odderon charge $\hat{q}_3$ \cite{jw2,korch3}.
Finally, in Ref.\cite{jw1} the singlevaluedness conditions for the
wave function $\varPhi$ were formulated and the spectrum of
$\Hat{q}_3$ was found. This allowed to calculate the energy
\cite{jw2} (hence also the odderon intercept) for the conformal
weight $h=1-\overline{h}=1/2$ which supposedly gives the largest
contribution to the elastic amplitude. It is, however, interesting
to see explicitly whether $h=1-\overline{h}=1/2$ gives really the
largest intercept and whether the there exist other solutions to
the singlevaluedness conditions than the ones found in \cite{jw1}.
The first attempt in this direction has been undertaken in
Ref.\cite{prasz} where the spectrum of $\Hat{q}_3$ for arbitrary
conformal weight but for non-physical quantization conditions
$h=\overline{h}$ was found. In Ref. \cite{br3} the spectrum of
$\Hat{q}_3$ for the specific choice of the conformal weight
$h=1-\overline{h}=1/2+i \nu$ in the limit of small $\nu$ has been
studied. This result of Ref. \cite{br3} was confirmed and extended
for arbitrary $\nu$ in Ref.\cite{prasz}.

The values of the conformal the weight $h$ and eigenvalues
of  $\Hat{q}_3$ form a four dimensional space.
In the present work we construct an algorithm and numerical code
which allows to find the points in the $(h,q_3)$ space
which satisfy the physical quantization conditions $h=1-\overline{h}$ and
singlevaluedness conditions of Refs.\cite{jw1,prasz}.
It turns out that these points form one dimensional curves in the
four dimensional space $(h,q_3)$. We have found families of curves which
are numbered by discrete values of $\RRe~h=1/2+m/2$ with $m \in 3 \mathbb Z$.
Therefore for given $\RRe~h$ these curves are effectively embedded in 3
dimensional subspace ($\IIm~h,\RRe~q_3,\IIm~q_3$).

Applying the method from
Ref.\cite{jw2} which allows to calculate the odderon energy for arbitrary
$h$ and $q_3$,  we have calculated the odderon energy
along the singlevaluedness lines $q_3(h)$. As expected, the odderon energy has
a maximum for $h=1-\overline{h}=1/2$,
and is always negative. Our numerical procedures are precise enough to find  17
values of $q_3$ for $h=1-\overline{h}=1/2$ with 9 digits accuracy.

The paper is organized as follows: in Section \ref{eqq3},
following Ref.\cite{prasz}, we write the odderon equation in terms
of variable $\xi$ suggested in Ref.\cite{Janik} and find its
solutions around $\xi= \pm 1$ and $\infty$. Recurrence relations
for the for these solutions are collected in Appendix A. Next, in
Sect.~\ref{quantq3}, we recapitulate the method of Ref.\cite{jw2}
and construct a singlevalued odderon wave function $\varPhi$
relegating the detailed form of the singlevaluedness constraints
to Appendix B. The resulting spectrum of $q_3$ and $h$ is
calculated and discussed in Sect.~\ref{sec:widq}. The numerical
algorithm used in this Section is described in detail in Appendix
C. Finally in Sect.~\ref{odden} we calculate the odderon energy
along the singlevaluedness curves found in Sect.~\ref{sec:widq}.
Conclusions are presented in Sect.~\ref{concl}.

\mysection{Solution of the eigenequation for the odderon charge $\hat{q}_3$}
\label{eqq3}

\subsection{Origin of the equation}

As already said in the Introduction it is possible to find  a family of
commuting operators $\Hat{q}_k$ which commute with the holomorphic $n$-Reggeon
Hamiltonian $\Hat{H}_n$ \cite{lip}:
\begin{equation}
[\Hat{H}_n,\Hat{q}_k]=0, \;\;\; k=2,\ldots,n.
\end{equation}
It follows that the Hamiltonian $\Hat{H}_n$ and the operators $\Hat{q}_k$
have the same set of eigenfunctions. In terms of the holomorphic coordinates
$\Hat{q}_k$ have the following form:
\begin{equation}
\Hat{q}_k=\sum_{n\geq i_1>i_2>\cdots>i_k\geq 1} i^k z_{i_1 i_2} z_{i_2 i_3}
\ldots  z_{i_k i_1} \partial_{i_1} \partial_{i_2} \ldots  \partial_{i_k},
\end{equation}
where  $k=2,\ldots,n$, $z_{jk}\equiv z_j-z_k$ and
$\partial_{j} \equiv \partial_{z_j} $.

For the odderon case, $n=3$, we have only 2 operators
\begin{equation}
\begin{aligned}
\Hat{q}_2&=\sum_{n\geq j>k\geq 1} z_{jk}^2 \partial_j \partial_k,\\
\Hat{q}_3&=\sum_{n\geq j>k>l\geq 1} -i z_{jk} z_{kl} z_{lj}
\partial_j \partial_k \partial_l .
\end{aligned}
\end{equation}

Following Ref.\cite{lip} we will use conformally covariant Ansatz for $\varPsi$
\begin{equation}
\varPsi(z_1,z_2,z_3)=z^{h/3} \psi(x),
\label{eq:psizx}
\end{equation}
where
\begin{equation}
z=\frac{(z_1-z_2)(z_1-z_3)(z_2-z_3)}{(z_1-z_0)^2(z_2-z_0)^2(z_3-z_0)^2},\;\;\;\;
x=\frac{(z_1-z_2)(z_3-z_0)}{(z_1-z_0)(z_3-z_2)},
\end{equation}
$h$ is a conformal weight and $z_0$ represents an arbitrary reference point.
A particular feature of this Ansatz (\ref{eq:psizx}) is that $\Hat{q}_2$
is automatically diagonal
\begin{equation}
\Hat{q}_2 \varPsi(z_1,z_2,z_3)=-h(h-1)\varPsi(z_1,z_2,z_3).
\end{equation}
In representation (\ref{eq:psizx}) the eigenvalue equation for $\Hat{q}_3$
takes the following form
\begin{equation}
\label{eq:jw}
\begin{aligned}
i\Hat{q}_3 \psi(x)=&\left(\frac{h}{3}\right)^2\left(\frac{h}{3}-1\right)
\frac{(x-2)(x+1)(2x-1)}{x(x-1)}\psi(x) \\
&+ \left[2x(x-1)-\frac{h}{3}(h-1)(x^2-x+1)\right]\psi'(x) \\
&+2x(x-1)(2x-1)\psi''(x)+x^2(x-1)^2\psi'''(x)
= i q_3 \psi(x).
\end{aligned}
\end{equation}

Equation (\ref{eq:jw}) has been studied in Ref.\cite{jw1} where
the quantization conditions for $\hat{q}_3$ were found by introducing the
singlevaluedness constraints on the whole wave function of odderon $\varPhi$.
The singlevaluedness conditions are much simpler when we
rewrite equation (\ref{eq:jw}) in terms of a new variable
suggested in Ref.\cite{Janik}
\begin{equation}
\label{eq:xi}
\xi=i \frac{1}{3 \sqrt{3}} \frac{(x-2)(x+1)(2x-1)}{x(x-1)}.
\end{equation}
Putting (\ref{eq:xi}) into (\ref{eq:jw}) we have \cite{prasz}
\begin{equation}
\label{eq:oddxi}
\begin{aligned}
\left[\frac{1}{2}(\xi^2-1)^2 \frac{d^3}{d\xi^3}+2\xi(\xi^2-1)
\frac{d^2}{d\xi^2}+\left(\frac{4}{9}-\beta_h (\xi^2-1)\right)\frac{d}{d\xi}
+\rho_h \xi +\tilde{q}\right]\varphi(\xi)=0,
\end{aligned}
\end{equation}
where
\[
\beta_h=\frac{(h+2)(h-3)}{6}, \ \ \ \rho_h=\frac{h^2(h-3)}{27},
\ \ \ \tilde{q}=\frac{q_3}{3\sqrt{3}},
\]
and $q_3$ is the eigenvalue of the operator $\Hat{q}_3$.

As we shall shortly see the odderon equation (\ref{eq:oddxi}) is less
singular than Eq.(\ref{eq:jw}) and the solutions of the indicial equation
around $\xi=\pm 1$ do not depend on $h$.

\subsection{Solution of the odderon equation}

The odderon equation (\ref{eq:oddxi}) has three regular singular points at
$\xi=-1$, $\xi=1$ and $\xi=\infty$.
We shall solve this equation using the power series method.
It is a third order ordinary differential equation therefore it has
three linearly independent solutions. We can write them as a vector
\begin{equation}
\Vec{u}(\xi;q_3)=
\left[
\begin{gathered}
u_1(\xi;q_3) \\
u_2(\xi;q_3) \\
u_3(\xi;q_3) \\
\end{gathered}
\right].
\label{eq:uxi}
\end{equation}

\subsubsection{Solution of the equation around $\xi=\pm 1$}

Solutions of equation (\ref{eq:oddxi}) around $\xi=\pm 1$ have
the following form \cite{prasz}:
\begin{equation}
\label{eq:upm}
u_i^{(\pm 1)}(\xi;q_3)=
(1\mp \xi)^{s_i} \sum_{n=0}^{\infty}u_{i,n}^{(\pm 1)}(\xi\mp 1)^n,
\end{equation}
where $s_i$ are solutions of the indicial equation
\begin{equation}
\label{eq:si}
s_1=\frac{2}{3}, \;\;\; s_2=\frac{1}{3}, \;\;\; s_3=0
\end{equation}
and do not depend on $h$.
The coefficients $u_{i,n}^{(\pm 1)}$ are defined in Appendix A.

\subsubsection{Solutions  around $\xi=\infty$}

The solution of equation (\ref{eq:oddxi}) around $\xi=\infty$ has
a more complicated form. In this case we perform a substitution
$\xi=1/\eta$ and then solve the problem around $\eta=0$:
\begin{equation}
\label{eq:oddeta}
\begin{aligned}
-&\frac{1}{2}(1-\eta^2)^2 \eta^2 \frac{d^3u}{d\eta^3} +
\left[2(1-\eta^2)\eta-3 \eta (1-\eta^2)^2\right] \frac{d^2u}{d\eta^2}\\
+&\left[-3(1-\eta^2)^2-\frac{4}{9} \eta^2 + (4+\beta_h) (1-\eta^2) \right]
 \frac{du}{d\eta}+\left[\rho_h \frac{1}{\eta}+ \tilde{q}\right]u =0
\end{aligned}.
\end{equation}
The solutions of the indicial equation $r_i$ depend on the conformal weight $h$
\begin{equation}
r_1=\frac{2h}{3}, \; \; \;  r_2=1-\frac{h}{3}, \; \; \; r_3=-\frac{h}{3}.
\end{equation}
and are identical as the solutions of the indicial equation in the
case of equation (\ref{eq:jw}). Since $r_2-r_3$ is equal to an
integer number, one of the solutions, $u_3^{(\infty)}(\xi)$,
contains a logarithm. The other two differences $r_2-r_1$,
$r_1-r_3$ become integer as well if $h$ is an integer itself.
Therefore we have to distinguish several cases.

For $h \notin \mathbb Z$ and $q_3 \neq 0$
the solution of  (\ref{eq:oddxi}) in vicinity of $\xi=\infty$
reads:
\begin{equation}
\label{eq:uinf}
\begin{aligned}
u_1^{(\infty)}(\xi;q_3)=&
(1/\xi)^{r_1} \sum_{n=0}^{\infty}u_{1,n}^{(\infty)}(1/\xi)^n \\
u_2^{(\infty)}(\xi;q_3)=&
(1/\xi)^{r_2} \sum_{n=0}^{\infty}u_{2,n}^{(\infty)}(1/\xi)^n \\
u_3^{(\infty)}(\xi;q_3)=&
(1/\xi)^{r_3} \sum_{n=0}^{\infty}u_{3,n}^{(\infty)}(1/\xi)^n
+u_2^{(\infty)}(\xi;q_3) \Log(1/\xi)
\end{aligned}
\end{equation}
where the logarithm $ \Log(z)$ is defined as:
\begin{equation}
\Log(z)=\ln|z|+i \Arg(z), \;\;\; |\Arg(z)|<\pi.
\end{equation}
The coefficients entering (\ref{eq:uinf}) are collected in
Appendix A.

The remaining cases, i.e. when $q_3=0$ and/or $h\in \mathbb Z$
\footnote {Solutions of the equation (\ref{eq:oddeta}) around $\xi=\infty$
for $h \in \mathbb Z$ are not considered in this work.}, should be considered
separately.

\subsubsection{Solutions around $\xi=\infty$ for $q_3=0$}

It is easy to observe that in equation (\ref{eq:uinfn}) for $q_3=0$
the term $u_{3,0}^{(\infty)}=\frac{1-h}{2\tilde{q}}$ tends to infinity.
In this case the solution of equation (\ref{eq:oddxi}) should be
constructed separately.
With $q_3=0$ and $h\notin \mathbb Z$ the solution is given by
\begin{equation}
\label{eq:uinfq0}
u_i^{(\infty)}(\xi;q_3=0)=
(1/\xi)^{r_i} \sum_{n=0}^{\infty}u_{i,2n}^{(\infty;q_3=0)}(1/\xi)^{2n}.
\end{equation}
The coefficients $u_{i,2n}^{(\infty;q_3=0)}$ are defined in Appendix A.

\subsection{Antiholomorphic sector}

For given $h$ and $q_3$ we can find the solutions of
Eq.(\ref{eq:oddxi}) around all singular points by means of
Eqs.(\ref{eq:upm}), (\ref{eq:uinf}) and (\ref{eq:uinfq0}).
Analogously, we can construct the solutions in the antiholomorphic
sector. Here instead of using the conformal weight $h$ and charge
$q_3$ we use their antiholomorphic equivalents: $\overline{h}$ and
$\overline{q}_3$\footnote{ {\em bar} does not denote complex
conjugation for which we us an {\em asterisk}.}. Similarly to
Eq.(\ref{eq:uxi}) we write the three linearly independent
solutions as a vector:
\begin{equation}
\Vec{v}(\xi^{\ast};\overline{q}_3)=
\left[
\begin{gathered}
v_1(\xi^{\ast};\overline{q}_3) \\
v_2(\xi^{\ast};\overline{q}_3) \\
v_3(\xi^{\ast};\overline{q}_3) \\
\end{gathered}
\right].
\end{equation}

\mysection{Quantization conditions for the odderon charge $\hat{q}_3$}
\label{quantq3}
\subsection{Transition matrices}

Each of the solutions around  $\xi = \xi_{1,2,3}$ where
$\xi_{1,2,3}=\pm 1,\infty$ has a convergence radius equal to the
difference between two singular points: the point around which the
solution is defined and the nearest of the remaining singular
points. In order to define the global solution which is convergent
in the entire complex plane we have to glue the solutions defined
around different singular points. This can be done by expanding
one set of solutions defined around $\xi_i$ in terms of the
solutions defined around $\xi_j$ for $\xi$ belonging to the
overlap region of the two solutions considered. Thus, in the
overlap region we can define the transition matrices $\varDelta$,
$\varGamma$, $\varOmega$, where
\begin{equation}
\label{eq:dgo}
\begin{aligned}
\Vec{u}^{(\infty)}(\xi;q_3)&=\varDelta(q_3)\Vec{u}^{(-1)}(\xi;q_3), \\
\Vec{u}^{(-1)}(\xi;q_3)&=\varGamma(q_3)\Vec{u}^{(+1)}(\xi;q_3),  \\
\Vec{u}^{(+1)}(\xi;q_3)&=\varOmega(q_3)\Vec{u}^{(\infty)}(\xi;q_3).
\end{aligned}
\end{equation}

Matrices $\varDelta$, $\varGamma$ and $\varOmega$ are constructed
in terms of the ratios of certain determinants. For example
to calculate the matrix $\varGamma$ we construct the Wro\'nskian
\begin{equation}
\label{eq:wron}
W=
\begin{vmatrix}
u_1^{(+1)}(\xi;q_3)& u_2^{(+1)}(\xi;q_3)& u_3^{(+1)}(\xi;q_3) \\
{u'}_1^{(+1)}(\xi;q_3)& {u'}_2^{(+1)}(\xi;q_3)& {u'}_3^{(+1)}(\xi;q_3) \\
{u''}_1^{(+1)}(\xi;q_3)& {u''}_2^{(+1)}(\xi;q_3)& {u''}_3^{(+1)}(\xi;q_3)
\end{vmatrix}.
\end{equation}
Next, we construct determinants $W_{ij}$ which are obtained from
$W$ by replacing  $j$-th column by the $i$-th solution around
$\xi=-1$, i.e. for $i=1$ and $j=2$ we have
\begin{equation}
W_{12}=
\begin{vmatrix}
u_1^{(+1)}(\xi;q_3)& u_1^{(-1)}(\xi;q_3)& u_3^{(+1)}(\xi;q_3) \\
{u'}_1^{(+1)}(\xi;q_3)& {u'}_1^{(-1)}(\xi;q_3)& {u'}_3^{(+1)}(\xi;q_3) \\
{u''}_1^{(+1)}(\xi;q_3)& {u''}_1^{(-1)}(\xi;q_3)& {u''}_3^{(+1)}(\xi;q_3)
\end{vmatrix}.
\end{equation}
The matrix elements $\varGamma_{ij}$ are defined as
\begin{equation}
\varGamma_{ij}=\frac{W_{ij}}{W}.
\end{equation}

Matrix $\varGamma$ does not depend on $\xi$, but only on $q_3$ and $h$.
In a similar way we can get the matrices $\varDelta$ and $\varOmega$
and their antiholomorphic equivalents:
$\overline{\varDelta}$, $\overline{\varGamma}$, $\overline{\varOmega}$.

\subsection{\label{sub:wkj}
Quantization conditions and singlevaluedness of the wave function}

The odderon charge $q_3$ is connected to its antiholomorphic equivalent
by
\begin{equation}
\label{eq:qqbar}
\overline{q}_3=-q_3^{\ast},
\end{equation}
where an {\em asterisk} over $q_3$ denotes  complex conjugation.
There exist two possible choices for $\overline{q}_3$: the one
given by Eq.(\ref{eq:qqbar}) and a similar one with the {\em plus} sign.
This follows from the fact that the eigenvalues of holomorphic and
antiholomorphic Hamiltonian, $\varepsilon_3$ and $\overline{\varepsilon}_3$,
are symmetric functions of $q_3$ and  $\overline{q}_3$ respectively \cite{korch}.
Only one choice, namely (\ref{eq:qqbar}), leads to the nonvanishing solution
of the quantization conditions\footnote{Note, that because of the factor $i$ in
the definition $\xi$ (\ref{eq:xi}), our $\overline{q}_3$ has different sign
than the one in Ref.\cite{jw1}.}.

The odderon wave function can be written as \cite{prasz}
\begin{equation}
\label{eq:phizxi}
\varPhi_{h\overline{h}q_3\overline{q}_3}(z,z^{\ast})=
z^{h/3}(z^{\ast})^{\overline{h}/3}
\Vec{v}^T(\xi^{\ast};\overline{q}_3)A(\overline{q}_3,q_3)\Vec{u}(\xi;q_3)
\end{equation}
where $\xi=\xi(z)$. The wave function
$\varPhi_{h\overline{h}q_3\overline{q}_3}(z,z^{\ast})$ contains
the solutions of equation (\ref{eq:oddxi}) and its antiholomorphic
counterpart, $\Vec{u}(\xi;q_3)$ and
$\Vec{v}(\xi^{\ast};\overline{q}_3)$ respectively, and a $3\times
3$ matrix $A(\overline{q}_3;q_3)$ "sewing" the solutions of the
both sectors.

The wave function $\varPhi$ has to be singlevalued. This means that
it should not depend on the choice of the Riemann sheet
for the variables $z$ and $\xi$.
In formula (\ref{eq:phizxi}) the term $z^{h/3}(z^{\ast})^{\overline{h}/3}$
is uniquely defined only if $h/3-\overline{h}/3 \in \mathbb Z$.
This leads to the quantization condition for the conformal weight $h$
\begin{equation}
\label{eq:hhbar}
h=\frac{1}{2}(\mu+m)+i\nu \;\;\; \mbox{  and  } \;\;\;
\overline{h}=\frac{1}{2}(\mu-m)+i\nu,
\end{equation}
where $\mu$ and $\nu$ are real and $m/3 \in \mathbb Z$.
The latter condition follows from the invariance under the Lorentz spin
transformations.
The normalization condition of the wave function requires that
$\mu=1$ for the physical odderon solution.

The fact that the wave function $\varPhi$ should be singlevalued
imposes certain conditions on the form of matrix $A$. It follows
from Eq.(\ref{eq:phizxi}) that for the solutions (\ref{eq:upm}) around
$\xi=\pm 1$ the matrix element $A_{ji}$ is multiplied by a factor
$(1\mp\xi)^{s_i}(1\mp\xi^{\ast})^{s_j}$.
This expression is singlevalued only if $s_i-s_j \in \mathbb Z$.
For $s_i$ of Eq.(\ref{eq:si}) this is true only for $i=j$. Therefore
the matrices $A^{(\pm 1)}$ have a diagonal form
\begin{equation}
A^{(-1)}(\overline{q}_3,q_3)=
\begin{bmatrix}
\alpha & 0 & 0 \\
0 & \beta & 0 \\
0 & 0 & \gamma \\
\end{bmatrix}, \;\;\;
A^{(+1)}(\overline{q}_3,q_3)=
\begin{bmatrix}
\alpha' & 0 & 0 \\
0 & \beta' & 0 \\
0 & 0 & \gamma' \\
\end{bmatrix}.
\label{eq:Apm}
\end{equation}

For the solutions around $\xi=\infty$ (\ref{eq:uinf},
\ref{eq:uinfq0}), and for $h\notin {\mathbb Z}$, the matrix
element $A_{ji}$ is multiplied by a factor
$(1/\xi)^{r_i}(1/\xi^{\ast})^{\overline{r}_j}$. One should notice
that solutions of the indicial equation $r_2$, $\overline{r}_2$,
$r_3$ and $\overline{r}_3$ differ by an
integer\footnote{$\overline{r}_2$ and $\overline{r}_3$ are
solutions  of the indicial equation in the antiholomorphic
sector.}, therefore terms which correspond to the elements
$A_{23}$ i~$A_{32}$, do not vanish. Furthermore, terms with a
logarithm appear  in the solutions $u_3^{(\infty)}$ and
$v_3^{(\infty)}$ for $q_3 \notin \mathbb Z$. One can see that when
$A_{23}=A_{32}$ then in the sum the ambiguous arguments of the
logarithms cancel out. Moreover, for $q_3 \notin \mathbb Z$ the
term which corresponds to the matrix element $A_{33}$ is not
singlevalued. It contains a square of the logarithm which does not
occur in any other terms. For this reason the element
$A_{33}(q_3\ne0,\overline{q}_3\ne0)$ has to vanish.

Thus the matrices $A$, defined around $\xi=\infty$,
have the following form
\begin{equation}
A^{(\infty)}(\overline{q}_3\ne 0,q_3 \ne 0)=
\begin{bmatrix}
\rho & 0 & 0 \\
0 & \sigma & \tau \\
0 & \tau & 0   \\
\end{bmatrix}, \;\;\;
A^{(\infty)}(\overline{q}_3=0,q_3=0)=
\begin{bmatrix}
\rho' & 0 & 0 \\
0 & \sigma' & \upsilon' \\
0 & \varsigma' & \tau' \\
\end{bmatrix}.
\label{eq:Ainf}
\end{equation}

Substituting  equation (\ref{eq:dgo}) into the wave function
(\ref{eq:phizxi}), one finds the following conditions for
the matrices $A(\overline{q}_3,q_3)$
\begin{eqnarray}
\label{eq:ainfam}
\overline{\varDelta}^T(\overline{q}_3)A^{(\infty)}(\overline{q}_3,q_3)\varDelta(q_3)&=
A^{(-1)}(\overline{q}_3,q_3), \\
\label{eq:amap}
\overline{\varGamma}^T(\overline{q}_3)A^{(-1)}(\overline{q}_3,q_3)\varGamma(q_3)&=
A^{(+1)}(\overline{q}_3,q_3), \\
\label{eq:apainf}
\overline{\varOmega}^T(\overline{q}_3)A^{(+1)}(\overline{q}_3,q_3)\varOmega(q_3)&=
A^{(\infty)}(\overline{q}_3,q_3).
\end{eqnarray}
Each of the formulae (\ref{eq:ainfam}-\ref{eq:apainf})
is equivalent to a set of nine equations which can be
conveniently written in terms of the following 4 vectors:
\[
\Vec{a}=
\begin{bmatrix}
\alpha \\
\beta \\
\gamma \\
\end{bmatrix},
 \;\;\;
\Vec{b}=
\begin{bmatrix}
\alpha' \\
\beta'\\
\gamma' \\
\end{bmatrix},
 \;\;\;
\Vec{c}=
\begin{bmatrix}
\rho \\
\sigma \\
\tau \\
\end{bmatrix},
 \;\;\;
\Vec{d}=
\begin{bmatrix}
\rho' \\
\sigma' \\
\varsigma' \\
\upsilon' \\
\tau' \\
\end{bmatrix},
\]
We can now rewrite equations (\ref{eq:ainfam},\ref{eq:amap},\ref{eq:apainf})
in the following form:
\begin{itemize}
\item equation (\ref{eq:ainfam}):
\begin{itemize}
\item for $q_3\ne 0$ as
\begin{equation}
\label{eq:buld}
B_{up} \Vec{c}=0, \;\;\; B_{low}\Vec{c}=0, \;\;\; B_{diag}\Vec{c}=\Vec{a},
\end{equation}
\item for $q_3= 0$ as
\begin{equation}
\label{eq:bpuld}
B'_{up} \Vec{d}=0, \;\;\; B'_{low}\Vec{d}=0, \;\;\; B'_{diag}\Vec{d}=\Vec{a}.
\end{equation}
\end{itemize}
\item equation (\ref{eq:amap}) as
\begin{equation}
\label{eq:culd}
C_{up} \Vec{a}=0, \;\;\; C_{low}\Vec{a}=0, \;\;\; C_{diag}\Vec{a}=\Vec{b},
\end{equation}
\item equation (\ref{eq:apainf}):
\begin{itemize}
\item for $q_3\ne 0$ as
\begin{equation}
\label{eq:duld}
D_{up} \Vec{b}=0, \;\;\; D_{low}\Vec{b}=0, \;\;\; D_{diag}\Vec{b}=\Vec{c},
\end{equation}
\item for $q_3 = 0$ as
\begin{equation}
\label{eq:dpuld}
D'_{up} \Vec{b}=0, \;\;\; D'_{low}\Vec{b}=0, \;\;\; D'_{diag}\Vec{b}=\Vec{d}.
\end{equation}
\end{itemize}
\end{itemize}
Definitions of matrices $B$, $C$, $D \ldots$ are given in Appendix B.

Equations (\ref{eq:buld}-\ref{eq:dpuld}) have nonvanishing solutions
only if the determinants of matrices with subscripts {\em up} and {\em low}
are equal zero. Moreover there should exist the unique
solutions of these equations:
$\Vec{a}$, $\Vec{b}$, $\Vec{c}$, $\Vec{d}$ which depend only
on one free parameter which can be fixed by normalizing
the wave function $\varPhi$. As we shall show in the next
Sections these two requirements fix uniquely the "boundary
conditions" for the eigen equation of the operator $\Hat{q}_3$
and allow to calculate its spectrum.

\mysection{Spectrum of the odderon charge $\hat{q}_3$ \label{sec:widq}}
\subsection{\label{sub:wwqh05} Eigenvalues of $\hat{q}_3$ for $h=1/2$}

Let us first discuss physical solutions found in Ref.\cite{jw1}
which correspond to the conformal weight $h=1/2$ and $\RRe~q_3=0$.
In order to calculate the eigenvalues $q_3$ we have solved
Eqs.(\ref{eq:buld}-\ref{eq:dpuld}) requiring  vanishing of the
{\em up} and {\em low} matrix determinants. After that, we have
also checked the uniqueness of obtained solutions. The results,
also for unphysical values of $q_3$ with $\IIm~q_3=0$ are
displayed in Table \ref{tab:q3h05}. Entries labelled from $0$ to
$4$, 12 and 13 agree with the ones of Refs.\cite{prasz,jw1}, while
the remaining ones are new.

\begin{table}[h]
\label{Tb:1}
\begin{center}
\begin{tabular}{|c|c|c|c||c|c|}
\hline \hline
No. & $q_3$ & No. & $q_3$ & No. & $q_3$ \\ \hline
0 &   $0$ &             6 &  $68.600522343i$ & 12 &   $1.475327424$ \\
1 &   $0.205257506i$ &  7 & $109.214406900i$ & 13 &  $12.947047037$ \\
2 &   $2.343921063i$ &  8 & $163.296192765i$ & 14 &  $44.413830163$ \\
3 &   $8.326345902i$ &  9 & $232.769867177i$ & 15 & $105.872614615$ \\
4 &  $20.080496894i$ & 10 & $319.559416811i$ & 16 & $207.320706051$ \\
5 &  $39.530550304i$ & 11 & $425.588828106i$ & 17 & $358.755426678$ \\
\hline \hline
\end{tabular}
\end{center}
\caption{Eigenvalues of the odderon charge $q_3$ for $h=1/2$}
\label{tab:q3h05}
\end{table}

In fact the eigenvalues $q_3$ for $h=1/2$ form a discrete set of
points symmetrically distributed on the real and complex axis in
complex $q_3$ plane. Therefore in Table \ref{Tb:1} only a half of
the spectrum is displayed. It has been shown \cite{jw2,jw1} that
only the imaginary values of $q_3$ are relevant for the odderon
problem; real eigenvalues correspond to a wave function which is
not totaly symmetric under the exchange of the neighboring
Reggeons. There exists also one eigenvalue $q_3=0$ which does not
correspond to a normalizable solution \cite{korch}, see however
\cite{BLV}.

\subsection{Eigenvalues $\hat{q}_3$ for the arbitrary conformal weights $h$}

As seen from Eq.(\ref{eq:hhbar}) the conformal weight $h$ depends
on two parameters: $m/3\in {\mathbb Z}$ and $\nu\in {\mathbb R}$.
In order to find  the manifolds on which conditions
(\ref{eq:buld}-\ref{eq:dpuld}) are satisfied we have extended the
domain of $h$ and allowed $m$ to be any real number. This defines
a four dimensional space of the conformal weight and the odderon
charge $(h,q_3) \in {\mathbb R}^4$. In each group of equations
(\ref{eq:buld}-\ref{eq:dpuld}) there are two equations which
contain matrices with subscripts {\em up} and {\em low}. The
determinants of these two matrices are complex, so we can define
the following fourvalued functions:
\begin{multline}
\label{eq:fb}
f_B : (\RRe~h,\IIm~h,\RRe~q_3,\IIm~q_3) \longrightarrow \\
\begin{cases}
\left(\RRe(\det B_{up}),\IIm(\det B_{up}),
\RRe(\det B_{low}),\IIm(\det B_{low})\right)
& {\rm for } \;\; q_3\ne0 \\
\left(\RRe(\det B'_{up}),\IIm(\det B'_{up}),
\RRe(\det B'_{low}),\IIm(\det B'_{low})\right)
& {\rm for } \;\; q_3=0 \\
\end{cases},\\
\end{multline}
\begin{multline}
\label{eq:fc}
f_C : (\RRe~h,\IIm~h,\RRe~q_3,\IIm~q_3) \longrightarrow \\
 (\RRe(\det C_{up}),\IIm(\det C_{up}),
 \RRe(\det C_{low} ),\IIm(\det C_{low} )),\\
\end{multline}
\begin{multline}
\label{eq:fd}
f_D : (\RRe~h,\IIm~h,\RRe~q_3,\IIm~q_3) \longrightarrow \\
\begin{cases}
\left(\RRe(\det D_{up}),\IIm(\det D_{up}),
\RRe(\det D_{low}),\IIm(\det D_{low})\right)
& {\rm for } \;\; q_3\ne0 \\
\left(\RRe(\det D'_{up}),\IIm(\det D'_{up}),
\RRe(\det D'_{low}),\IIm(\det D'_{low})\right)
& {\rm for } \;\; q_3=0 \\
\end{cases}.
\end{multline}

Thus, in order to calculate the spectrum of the operator $\Hat{q}_3$,
we should find common zeros of all functions $f_B$, $f_C$ and $f_D$:
\begin{equation}
f_B=0, ~~~~f_C=0, ~~~~f_D=0.
\label{eq:zeros}
\end{equation}
Furthermore one should verify the uniqueness of the solutions for
(\ref{eq:buld}-\ref{eq:dpuld}).

In Appendix C we have described the numerical algorithm constructed
to find roots of Eqs.(\ref{eq:zeros}). Our numerical findings
can be summarized as follows:
\begin{enumerate}
\item Although we have formally allowed $m$ to be a continuous real
parameter, the solutions of Eqs.(\ref{eq:zeros}) exist only for
$m/3 \in \mathbb Z$ ({\em e.g.} $\RRe~h=1/2$ or $\RRe~h=2$).
\item For the above discrete values of $m$, that is for fixed
$\RRe~h$, the solutions of Eqs.(\ref{eq:zeros}) form continuous
curves in 3 dimensional subspace $(\IIm~h,\RRe~q_3,\IIm~q_3)$.
\item It turned out that each of 3 equations (\ref{eq:zeros})
yields the same set of curves, provided that the solutions of
Eqs.(\ref{eq:buld}-\ref{eq:dpuld}) are unique.
\end{enumerate}

In the following we discuss two sets of solutions to
Eqs.(\ref{eq:zeros}), namely for $\RRe~h=1/2$ and $\RRe~h=2$.

\unitlength 1mm
\begin{figure}[h]
\epsfxsize=14cm
\frame{\epsffile{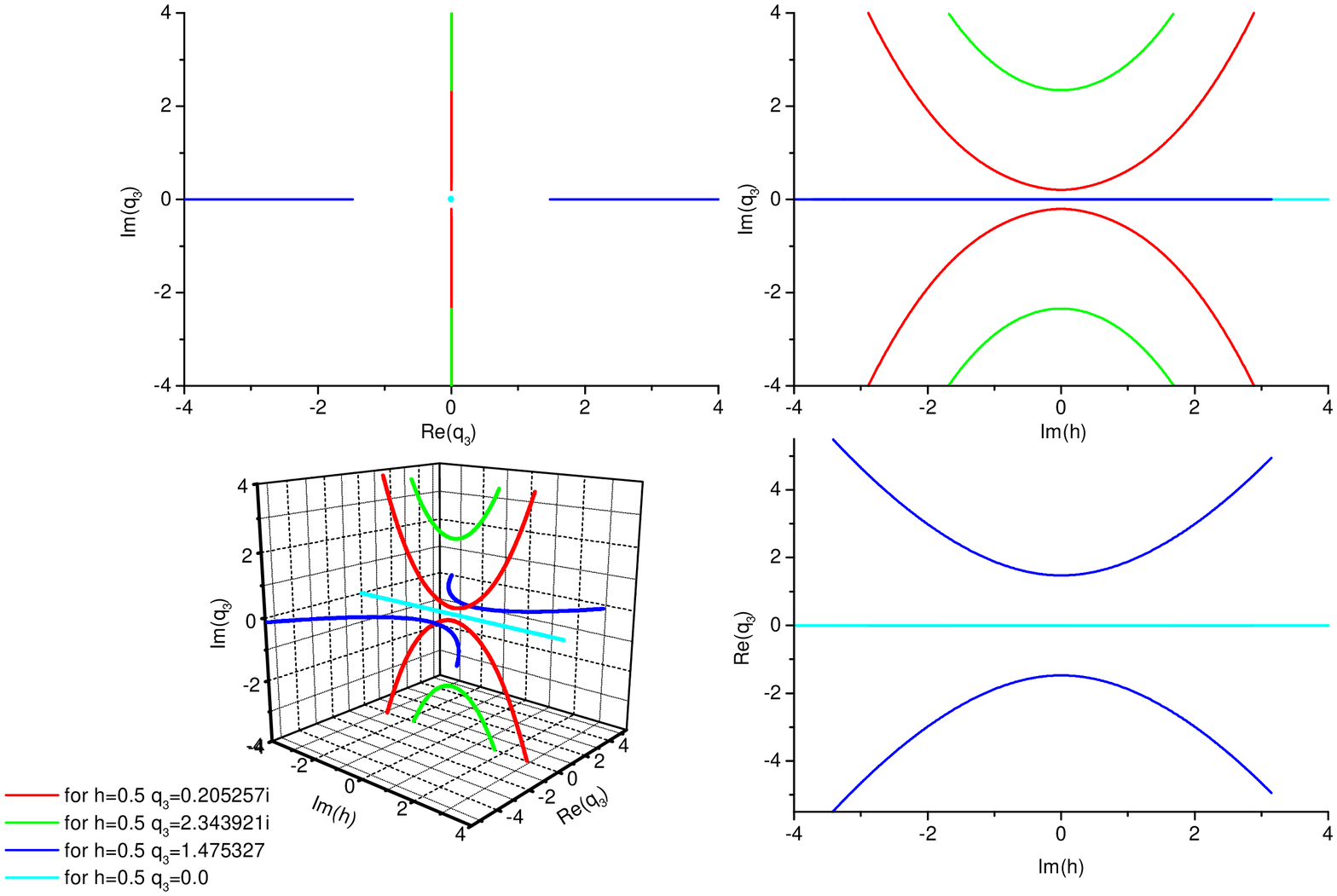}}
\caption[Odderon charge $q_3$ as a function of $h$ for $\RRe~h=1/2$]
{Odderon charge $q_3$ as a function of $h$ for $\RRe~h=1/2$
Presented curves are described with point with the higher value of $|q_3|$.}
\label{rys:plot4}
\end{figure}


\unitlength 1mm
\begin{figure}[h]
\epsfxsize=14cm
\frame{\epsffile{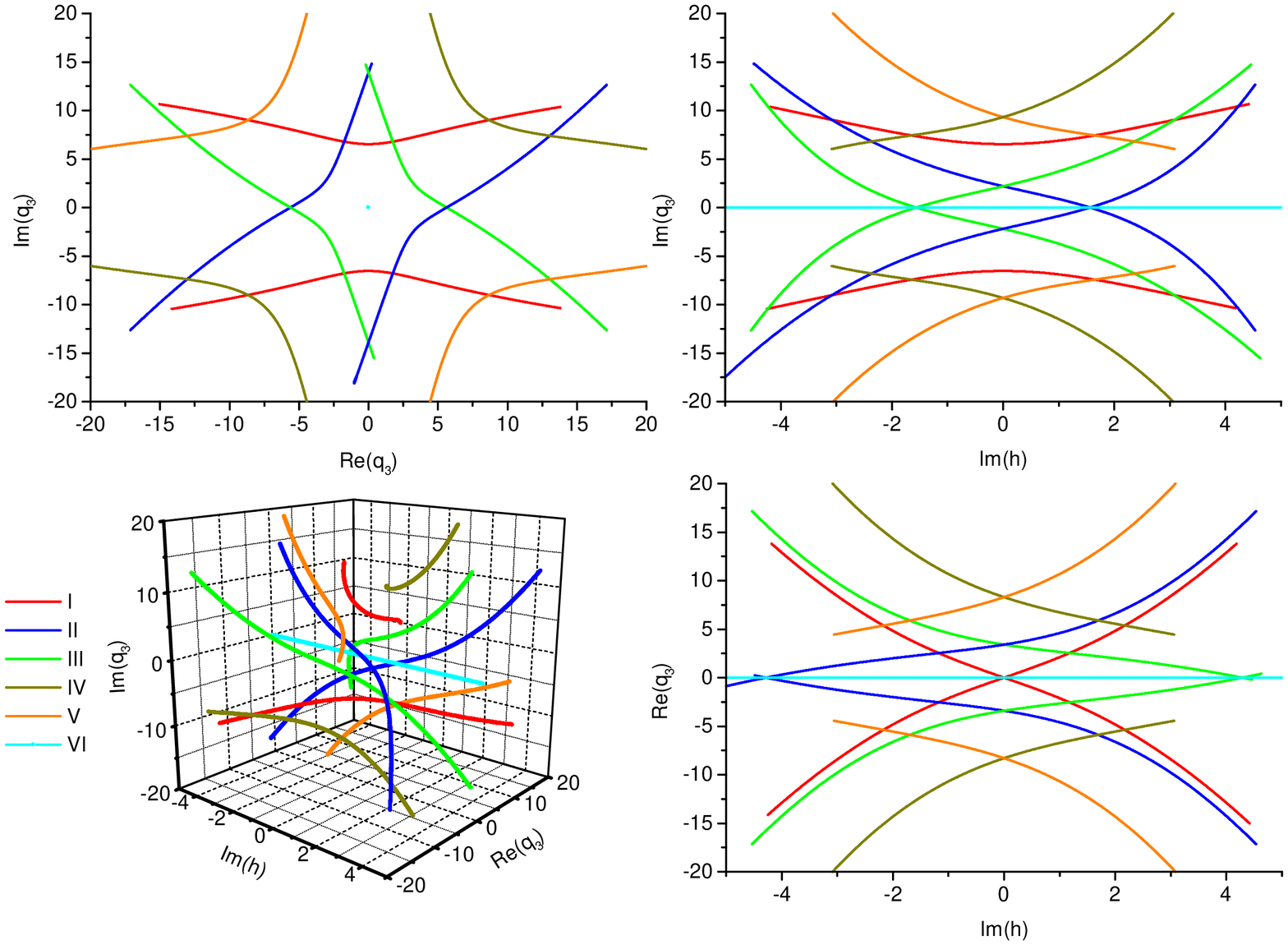}}
\caption{Odderon charge $q_3$ as a function of $h$ for $\RRe~h=2$}
\label{rys:plot5}
\end{figure}

\subsubsection{Spectrum of $\hat{q}_3$ for $Re(h)=1/2$}

In Figure \ref{rys:plot4} we plot spectrum of the odderon charge $\Hat{q}_3$
as a function of $h$ for $\RRe~h=1/2$. One can see two sets of curves: the
ones in the plane of $\RRe~q_3=0$ and in the perpendicular plane of
$\IIm~q_3=0$ and the line $q_3=0$ which belongs to the both classes.
For all these curves, except for $q_3=0$, the minimum of $|q_3|$ occurs
for $h=1/2$. Going away from this point the absolute value of $q_3$
increases monotonically. Minimal values of $|q_3|$ correspond to the
points listed in Table \ref{tab:q3h05}. We have not found any curve
located outside of $\RRe~q_3=0$ or $\IIm~q_3=0$ planes.

The curves in the plane of $\RRe~q_3=0$ have been earlier found
in Ref.\cite{braun}  and also in Refs.\cite{prasz,kw}.

The spectrum of $q_3$ has the following symmetries:
\begin{enumerate}
\item $\RRe~q_3 \rightarrow - \RRe~q_3$,
\item $\IIm~q_3 \rightarrow - \IIm~q_3$,
\item $h \rightarrow 1-h$
\end{enumerate}
which follow from the symmetries of the odderon equation
(\ref{eq:oddxi}). Indeed, equation (\ref{eq:oddxi}) is invariant
under the transformation $\xi\rightarrow -\xi$ and $q_3\rightarrow -q_3$.
The last symmetry follows from the exchange symmetry between the holomorphic
and antiholomorphic sectors.

\unitlength 1mm
\begin{figure}[h]
\epsfxsize=14cm
\frame{\epsffile{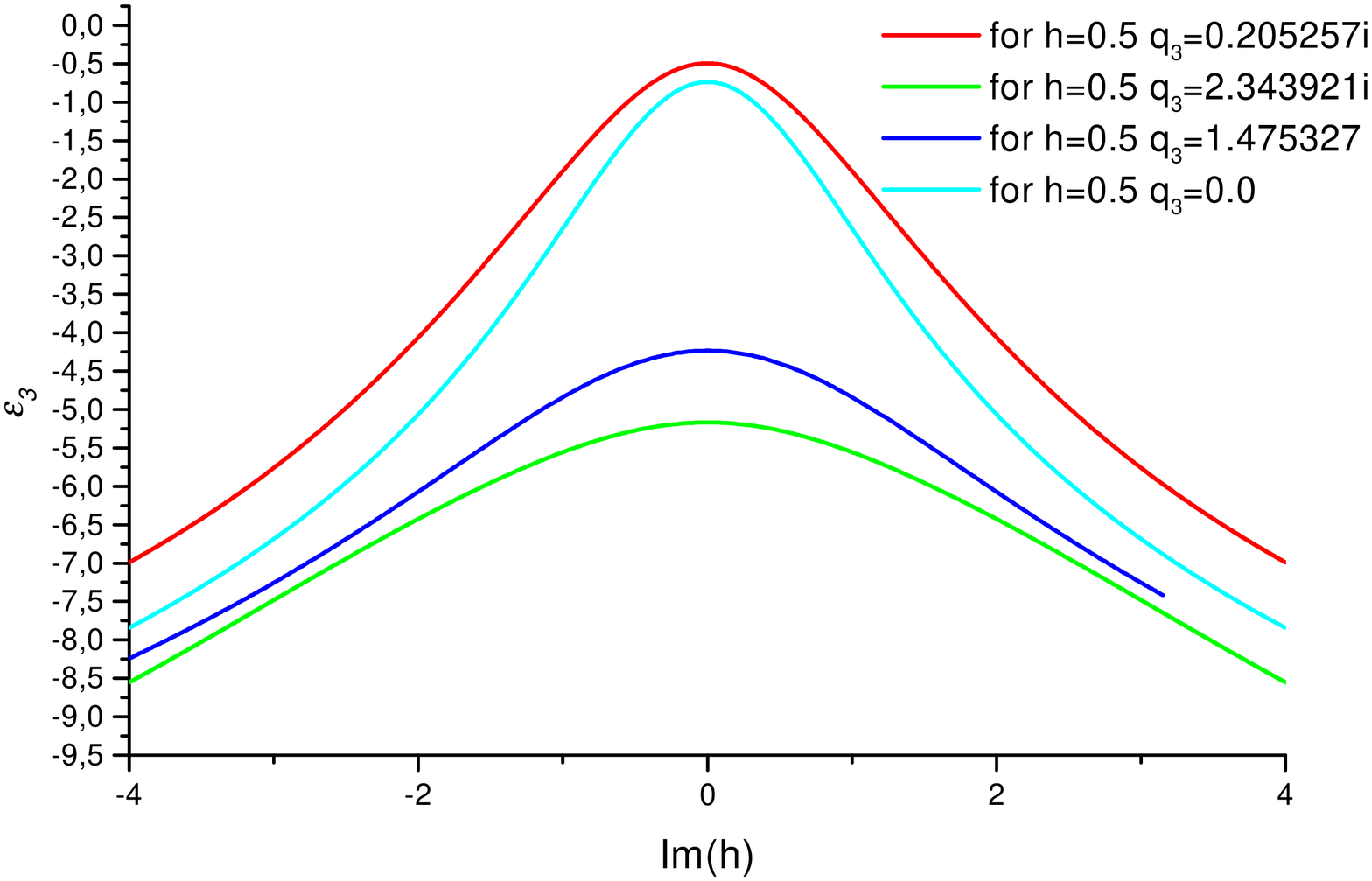}}
\caption[Real part of the holomorphic odderon energy for $\RRe~h=1/2$]
{Real part of holomorphic energy of odderon for $\RRe~h=1/2$.
Picture is plotted for curves from figure \ref{rys:plot4}.}
\label{rys:eplot4}
\end{figure}

\subsubsection{Spectrum of $\hat{q}_3$ for $Re(h)=2$}

In Figure \ref{rys:plot5} we plot the odderon charge $q_3$
as a function of the conformal weight for $\RRe~h=2$.
Except of a line with $q_3=0$ which is analogous to the case
of $\RRe~h=1/2$, the remaining curves have  more complicated
character. Still, they obey the following symmetries:
\begin{enumerate}
\item $q_3 \rightarrow - q_3$,
\item $\IIm~h \rightarrow - \IIm~h \;\;\;\RRe~q_3 \rightarrow - \RRe~q_3$,
\item $\IIm~h \rightarrow - \IIm~h \;\;\;\IIm~q_3 \rightarrow - \IIm~q_3$.
\end{enumerate}
These symmetries are similar to the case of $h=1/2$.

The first one is connected to Bose symmetry ($\xi \leftrightarrow -\xi$)
and the others
correspond to exchange of holomorphic and antyholomorphic sectors.
For $h=2$, that is, if the imaginary part of $h$ vanishes,
equation (\ref{eq:oddxi}) has not been solved. In this case the
eigenvalues $q_3$ have been obtained by an interpolation of the
neighboring points which satisfied the quantization conditions.

\subsubsection{Case for $q_3=0$ }

During the numerical computations we have noticed that similarly
to the other  matrices $A$, the matrix
$A^{(\infty)}(q_3=0,\overline{q}_3=0)$ depends only on three
parameters. The remaining ones vanish identically. In the case
$\RRe~h=1/2$, $\varsigma'$  and $\upsilon'$ vanish and for
$\RRe~h=2$, $\sigma'=\tau'=0$.

\mysection{Odderon energy} \label{odden}

The odderon energy is defined as \cite{korch}
\begin{equation}
E_3=\frac{\alpha_s N_c}{4 \pi}
\left[\varepsilon_3(h,q_3)+\bar{\varepsilon}_3(\bar{h},\bar{q}_3)\right]
=\frac{\alpha_s N_c}{2 \pi}\RRe (\varepsilon_3(h,q_3)),
\end{equation}
where $\varepsilon_3$  and $\bar{\varepsilon}_3$ are the largest eigenvalues
of the holomorphic and antiholomorphic odderon Hamiltonian respectively.
Applying the Bethe Ansatz we have for $n=3$ \cite{korch,lip2}
\begin{equation}
\varepsilon_3=
i \left(\frac{\dot{Q}_3(-i)}{Q_3(-i)}-\frac{\dot{Q}_3(i)}{Q_3(i)} \right)-6.
\end{equation}
where $Q_3(\lambda)$ satisfies the following Baxter equation
\cite{bax}:
\begin{equation}
\label{eq:baxter}
(\lambda+i)^3 Q_3(\lambda+i)+(\lambda-i)^3 Q_3(\lambda-i)
-(2 \lambda^3 +\lambda q_2+ q_3)Q_3(\lambda)=0.
\end{equation}
Equation (\ref{eq:baxter}) was solved in Ref.\cite{jw2} by a substitution
\begin{equation}
\lambda^k Q_k(\lambda)=\int_{C_z} \frac{dz}{2 \pi i} K(z,\lambda) \hat{P}^k Q(z)
-i \sum_{m=1}^{k}\int_{C_z} \frac{dz}{2 \pi i}
\dz{} \left[z(z-1)\hat{L}^{k-m} K(z,\lambda)\hat{P}^{m-1} Q(z) \right],
\end{equation}
where
\begin{equation}
K(z,\lambda)=z^{-i \lambda-1}(z-1)^{i \lambda-1}, \; \;
\hat{L}=\left(-i\dz{} z(z-1)\right), \; \;
\hat{P}=\left( i z(z-1) \dz{} \right).
\end{equation}
Choosing the proper integration contour $C_z$ and boundary conditions
one arrives at a differential equation for $Q(z)$:
\begin{equation}
\label{eq:rbaxter}
\left[\left(z(z-1)\dz{}\right)^3
- q_2 z^2(z-1)^2 \dz{}-i q_3 z(1-z) \right] Q(z)=0
\end{equation}
Similarly to the solutions of equation (\ref{eq:oddxi}), the
solutions of Eq.(\ref{eq:rbaxter}) depend on the conformal weight
$h$ and the odderon charge $q_3$. Using the spectrum $q_3(h)$
calculated in the last Section, we have calculated the energy of
the odderon along the curves from Figures \ref{rys:plot4} and
\ref{rys:plot5}.

Analyzing the spectra of energy we can conclude that the odderon
energy is always negative. This means that the intercept
\begin{equation}
\alpha (t=0)=E_3+1
\end{equation}
is lower than one, so the odderon partial amplitude $A_3(s,t)$
(\ref{eq:An}) is
described by the convergent series in Regge limit (\ref{eq:rlim}).

\subsection{Spectrum of the energy for $\RRe~h=1/2$}

\begin{table}[h]
\begin{center}
\begin{tabular}{|c|c|} \hline \hline
$q_3$       & $\RRe(\varepsilon_{3})$  \\ \hline
$0$         & $-0.73801$\\
$0.20526i$  & $-0.49434$\\
$2.34392i$  & $-5.16930$\\
$1.47533 $  & $-4.23462$\\
\hline \hline
\end{tabular}
\end{center}
\caption{Maximal values of $\RRe(\varepsilon_3)$ for $\RRe~h=1/2$}
\label{tab:eh05}
\end{table}

\unitlength 1mm
\begin{figure}[h]
\epsfxsize=14cm \frame{\epsffile{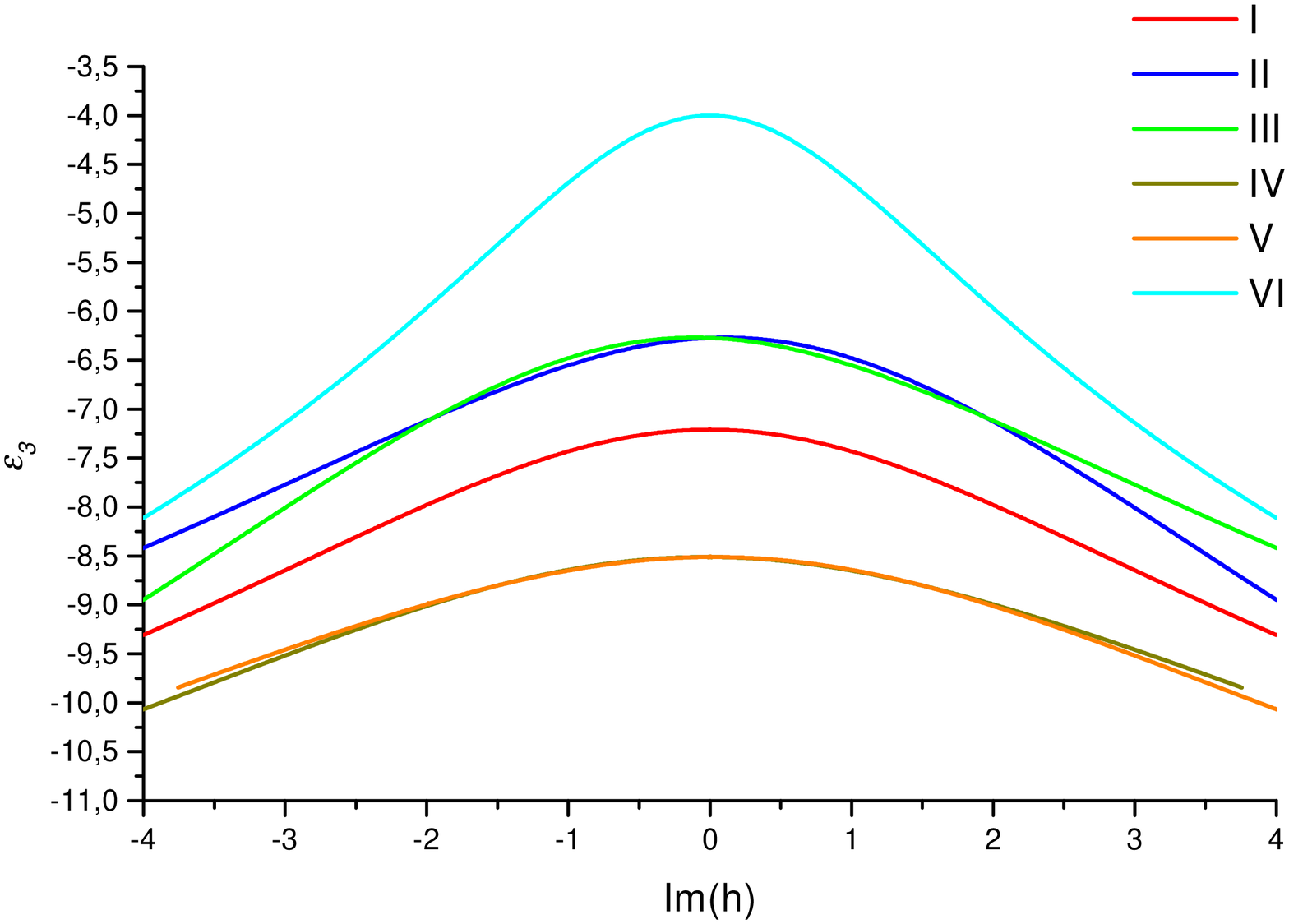}} \caption{Real part
of the holomorphic odderon energy for $\RRe~h=2$. The picture is
plotted for the curves from figure \ref{rys:plot5}.}
\label{rys:eplot5}
\end{figure}

In Figure \ref{rys:eplot4} we plot a real part of the holomorphic
odderon energy $\RRe(\varepsilon_3)$ as a function of $\IIm~h$.
The picture is plotted for curves from Figure \ref{rys:plot4}, that is
for $\RRe~h=1/2$.
All these curves have a maximum in $h=1/2$. The maximal values
are displayed in Table \ref{tab:eh05}.

Going away from the maximum, the energy decreases monotonically.
Our results agree with the values from Ref.\cite{jw1}.
The energy spectrum has the following symmetry
\begin{equation}
h\longrightarrow 1-h.
\end{equation}

\subsection{Spectrum of the energy for $\RRe~h=2$}

In Figure \ref{rys:eplot5}, similarly to Fig.~\ref{rys:eplot4},
we plot $\RRe(\varepsilon_3)$ as a function of $\IIm~h$ for
curves from Figure \ref{rys:plot5}, that is
for $\RRe~h=2$.
In this case the curves in $(h,q_3)$ space have  much more complicated
character and location of the energy maxima occurs not always for
$\IIm~h=0$.

\begin{table}[h]
\begin{center}
\begin{tabular}{|c|c|} \hline \hline
curve symbols & $\RRe(\varepsilon_{3})$  \\ \hline
I       & $-7.21$\\
II, III     & $-6.27$\\
IV, V       & $-8.51$\\
VI      & $-4.00$\\
\hline \hline
\end{tabular}
\end{center}
\caption{Maximal value $\RRe(\varepsilon_3)$ for $\RRe~h=2$}
\label{tab:eh20}
\end{table}

For plots I, VI, $\RRe(\varepsilon_3)$ has a maximum in $h=2$. In
others cases the energy has a maximum in vicinity of $h=2$, i.e.
for the II-nd curve the energy has a maximum in $(h=2.0+0.107i,
q_3=-3.508+2.050i)$. The maximal values of energy are given in
Table \ref{tab:eh20}. Similarly to the  case of $\RRe~h=1/2$ going
away from the maximum, energy decreases monotonically.

\mysection{Summary and Conclusions} \label{concl}

The aim of the present paper was to look for the solutions
of the odderon equation (\ref{eq:oddxi}) in the entire four
dimensional space of the conformal weight $h$ and odderon charge
$q_3$. So far these solutions have been found only for some
specific values of $h$ \cite{jw2,braun,kw,br3} or for arbitrary $h$
but unphysical quantization condition $h=\overline{h}$ \cite{prasz}.
To this end we have constructed and implemented the algorithm
which is in detail described in Appendix C. This algorithm
can be easily extended to more dimensional cases \cite{dkkm}.

The calculations were performed in the holomorphic variable $\xi$
which respects Bose symmetry and was proposed in Refs.\cite{Janik,prasz}.
The odderon equation (\ref{eq:oddxi}) is a third order ordinary
differential equation with three regular singular points at
$\xi=-1$, $1$ and $\infty$.
Solutions around $\xi=\pm1$ have been already found
in Ref.\cite{prasz}. Here we have also calculated solutions
around $\xi=\infty$ (\ref{eq:oddeta}).

The singlevaluedness conditions imposed on the odderon wave
function $\varPhi$ were found to be fulfilled along the discrete
sets  of continuous one dimensional curves. These sets are
numbered by values of $\RRe~h=1/2 + m/2$ ($m/3 \in \mathbb Z$) and
lie effectively in 3 dimensional subspace $(\IIm~h, \RRe~q_3,
\IIm~q_3)$. In this way we have obtained numerically the known
spectrum of a Lorentz spin $m$.

Although there are in principle 3 different singlevaluedness
conditions obtained by gluing solutions around each of 3 singular
points, which have to be fulfilled {\em simultaneously}, it turned
out that it was enough to satisfy only one of them to get a
complete set of solutions. It was therefore enough to consider two
singular points, namely $\pm 1$ for which the solutions of the
characteristic equation do not depend on $h$.

Finally, we have calculated the odderon energy along
the singlevaluedness curves $q_3(h)$. For all cases the
energy turned out to be negative which means that the odderon
intercept is smaller than 1. The maximal value of the odderon
energy corresponds, as earlier conjectured, to $h=1/2$ and
$q_3=0.205257506 \times i$ which can be seen on Fig.~3.

\vspace{1cm}

The authors thank J. Wosiek and G. Korchemsky for valuable comments
and discussion. We are grateful to J. Wosiek and A.~Rostworowski for
making the program for calculating the odderon energy available to us.
This work was partially supported by the Polish KBN Grant PB~2~P03B~{\-}%
019~17.

\newpage


\newpage

\appendix
\newcommand{\LI}{\mbox{{\em L1}}}
\newcommand{\LII}{\mbox{{\em L2}}}
\newcommand{\LIII}{\mbox{{\em L3}}}
\newcommand{\NI}{\mbox{{\em N1}}}
\newcommand{\NII}{\mbox{{\em N2}}}
\newcommand{\NIII}{\mbox{{\em N3}}}
\newcommand{\M}{\mbox{\em M}}
\renewcommand{\theequation}{\Alph{section}.\arabic{equation}}
\mysection{Definition of series from solutions of the equation for $\hat{q}_3$}
\subsection{Solutions around $\xi=\pm 1$}

The solutions of the equation (\ref{eq:oddxi}) around $\xi=\pm 1$
have a form (\ref{eq:upm}), where coefficients are defined by
\begin{equation}
\begin{aligned}
u_{i,0}^{(\pm 1)}=&1, \;\;\;\;
u_{i,1}^{(\pm 1)}=a_{i,0}
u_{i,0}^{(\pm 1)}/m_{i,1}, \\
u_{i,n}^{(\pm 1)}=&\left(a_{i,n-1} u_{i,n-1}^{(\pm 1)}
 +b_{i,n-2} u_{i,n-2}^{(\pm 1)}\right)/m_{i,n},
\end{aligned}
\end{equation}
while
\begin{equation}
\begin{aligned}
a_{i,n-1}=&\mp 4 (n+s_i-1)
\left[(n+s_i-2)(n+s_i)-\beta_h\right]+2\rho_h \pm 2\tilde{q}, \\
b_{i,n-2}=&-(n+s_i-2)
\left[(n+s_i-3)(n+s_i)-2\beta_h\right]+2\rho_h, \\
m_{i,n}=&4(n+s_i)\left[(n+s_i)(n+s_i-1)+2/9\right].
\end{aligned}
\end{equation}
The upper sign corresponds to solutions around $\xi=1$
and the lower one to solutions around $\xi=-1$.

\subsection{Solutions around $\xi=\infty$}

The coefficients of solutions of equation (\ref{eq:uinf}) for $q_3\ne 0$
and $h\notin \mathbb Z$ we can write as
\begin{itemize}
\item for $i=1,2$:
\begin{equation}
\begin{aligned}
u_{i,0}^{(\infty)}=&1, \;\;\;\;
u_{i,1}^{(\infty)}=a_{i,0}
u_{i,0}^{(\infty)}/m_{i,1}, \\
u_{i,2}^{(\infty)}=&\left(a_{i,1} u_{i,1}^{(\infty)}
 +b_{i,0} u_{i,0}^{(\infty)}\right)/m_{i,2},\\
u_{i,3}^{(\infty)}=&\left(a_{i,2} u_{i,2}^{(\infty)}
 +b_{i,1} u_{i,1}^{(\infty)}\right)/m_{i,3},\\
u_{i,n}^{(\infty)}=&\left(a_{i,n-1} u_{i,n-1}^{(\infty)}
 +b_{i,n-2} u_{i,n-2}^{(\infty)}
 +c_{i,n-4} u_{i,n-4}^{(\infty)}\right)/m_{i,n},
\end{aligned}
\end{equation}
\item for $i=3$:
\begin{equation}
\label{eq:uinfn}
\begin{aligned}
u_{3,0}^{(\infty)}=&\frac{1-h}{2\tilde{q}}, \;\;\;\;
u_{3,1}^{(\infty)}=0, \\
u_{3,2}^{(\infty)}=&\left(
  a_{3,1} u_{3,1}^{(\infty)}
 +b_{3,0} u_{3,0}^{(\infty)}
 +d_{3,1} u_{2,1}^{(\infty)}
 \right)/m_{i,2},\\
u_{3,3}^{(\infty)}=&\left(
  a_{3,2} u_{3,2}^{(\infty)}
 +b_{3,1} u_{3,1}^{(\infty)}
 +d_{3,2} u_{2,2}^{(\infty)}
 +f_{3,0} u_{2,0}^{(\infty)}
 \right)/m_{3,3},\\
u_{3,4}^{(\infty)}=&\left(
  a_{3,3} u_{3,3}^{(\infty)}
 +b_{3,2} u_{3,2}^{(\infty)}
 +c_{3,0} u_{3,0}^{(\infty)}
 +d_{3,3} u_{2,3}^{(\infty)}
 +f_{3,1} u_{2,1}^{(\infty)}
 \right)/m_{3,4},\\
u_{3,n}^{(\infty)}=&\left(
  a_{3,n-1} u_{3,n-1}^{(\infty)}
 +b_{3,n-2} u_{3,n-2}^{(\infty)}
 +c_{3,n-4} u_{3,n-4}^{(\infty)}
 \right. \\
 &\left.
+d_{3,n-1} u_{2,n-1}^{(\infty)}
 +f_{3,n-3} u_{2,n-3}^{(\infty)}
 +g_{3,n-5} u_{2,n-5}^{(\infty)}
 \right)/m_{i,n},
\end{aligned}
\end{equation}
where
\begin{equation}
\begin{aligned}
a_{i,n-1}=&2 \tilde{q},\\
b_{i,n-2}=&2(n+r_i-2)
\left[(n+r_i-1)(n+r_i-2)-\beta_h-4/9 \right],\\
c_{i,n-4}=&-(n+r_i-4)(n+r_i-3)(n+r_i-2),\\
d_{i,n-1}=&-(n+r_i)
\left[3(n+r_i-2)+4 \right]+2(1+\beta_h),\\
f_{i,n-3}=&2
\left[(n+r_i-2)(3(n+r_i-4)+8)-\beta_h-4/9 \right],\\
g_{i,n-5}=&-3(n+r_i-4)(n+r_i-2)-2,\\
m_{i,n}=&(n+r_i)^2(n+r_i-1)-2(n+r_i)(1+\beta_h)-2 \rho_h.
\end{aligned}
\end{equation}
\end{itemize}
Similarly the coefficients for the solution (\ref{eq:uinfq0}) for
$q_3=0$ and $h\notin \mathbb Z $ have a form
\begin{equation}
\begin{aligned}
u_{i,0}^{(\infty;q_3=0)}=&1, \;\;\;\;
u_{i,2}^{(\infty;q_3=0)}=b_{i,0}
u_{i,0}^{(\infty)}/m_{i,2}, \\
u_{i,2n}^{(\infty;q_3=0)}=&\left(
b_{i,2(n-1)} u_{i,2(n-1)}^{(\infty;q_3=0)}
+c_{i,2(n-2)} u_{i,2(n-2)}^{(\infty;q_3=0)}
\right)/m_{i,2n},\\
\end{aligned}
\end{equation}

\mysection{Definition of matrices connected to
singlevaluedness constraints on $\varPhi$}

In the section \ref{sub:wkj} we defined quantization conditions
for operator $\Hat{q}_3$. The matrices in formula (\ref{eq:buld})
have a following form
\begin{equation}
B_{up}=
\begin{bmatrix}
\overline{\varDelta}_{11} \varDelta_{12} &
\overline{\varDelta}_{21} \varDelta_{22} &
\overline{\varDelta}_{31} \varDelta_{22} + \overline{\varDelta}_{21} \varDelta_{32}  \\
\overline{\varDelta}_{11} \varDelta_{13} &
\overline{\varDelta}_{21} \varDelta_{23} &
\overline{\varDelta}_{31} \varDelta_{23} + \overline{\varDelta}_{21} \varDelta_{33}  \\
\overline{\varDelta}_{12} \varDelta_{13} &
\overline{\varDelta}_{22} \varDelta_{23} &
\overline{\varDelta}_{32} \varDelta_{23} + \overline{\varDelta}_{22} \varDelta_{33}  \\
\end{bmatrix} ,
\end{equation}
\begin{equation}
B_{low}=B_{up}(\varDelta \leftrightarrow \overline{\varDelta}),
\end{equation}
\begin{equation}
B_{diag}=
\begin{bmatrix}
\overline{\varDelta}_{11} \varDelta_{11} &
\overline{\varDelta}_{21} \varDelta_{21} &
\overline{\varDelta}_{31} \varDelta_{21} + \overline{\varDelta}_{21} \varDelta_{31}  \\
\overline{\varDelta}_{12} \varDelta_{12} &
\overline{\varDelta}_{22} \varDelta_{22} &
\overline{\varDelta}_{32} \varDelta_{22} + \overline{\varDelta}_{22} \varDelta_{32}  \\
\overline{\varDelta}_{13} \varDelta_{13} &
\overline{\varDelta}_{23} \varDelta_{23} &
\overline{\varDelta}_{33} \varDelta_{23} + \overline{\varDelta}_{23} \varDelta_{33}  \\
\end{bmatrix}.
\end{equation}
We can write matrices from (\ref{eq:bpuld}) as
\begin{equation}
B'_{up}=
\begin{bmatrix}
\overline{\varDelta}_{11} \varDelta_{12} &
\overline{\varDelta}_{21} \varDelta_{22} &
\overline{\varDelta}_{31} \varDelta_{22} &
\overline{\varDelta}_{21} \varDelta_{32} &
\overline{\varDelta}_{31} \varDelta_{32} \\
\overline{\varDelta}_{11} \varDelta_{13} &
\overline{\varDelta}_{21} \varDelta_{23} &
\overline{\varDelta}_{31} \varDelta_{23} &
\overline{\varDelta}_{21} \varDelta_{33} &
\overline{\varDelta}_{31} \varDelta_{33} \\
\overline{\varDelta}_{12} \varDelta_{13} &
\overline{\varDelta}_{22} \varDelta_{23} &
\overline{\varDelta}_{32} \varDelta_{23} &
\overline{\varDelta}_{22} \varDelta_{33} &
\overline{\varDelta}_{32} \varDelta_{33} \\
\overline{\varDelta}_{12} \varDelta_{11} &
\overline{\varDelta}_{22} \varDelta_{21} &
\overline{\varDelta}_{32} \varDelta_{21} &
\overline{\varDelta}_{22} \varDelta_{31} &
\overline{\varDelta}_{32} \varDelta_{31} \\
\overline{\varDelta}_{13} \varDelta_{11} &
\overline{\varDelta}_{23} \varDelta_{21} &
\overline{\varDelta}_{33} \varDelta_{21} &
\overline{\varDelta}_{23} \varDelta_{31} &
\overline{\varDelta}_{33} \varDelta_{31} \\
\end{bmatrix} ,
\end{equation}
\begin{equation}
B'_{low}=
\begin{bmatrix}
\overline{\varDelta}_{11} \varDelta_{12} &
\overline{\varDelta}_{21} \varDelta_{22} &
\overline{\varDelta}_{31} \varDelta_{22} &
\overline{\varDelta}_{21} \varDelta_{32} &
\overline{\varDelta}_{31} \varDelta_{32} \\
\overline{\varDelta}_{11} \varDelta_{13} &
\overline{\varDelta}_{21} \varDelta_{23} &
\overline{\varDelta}_{31} \varDelta_{23} &
\overline{\varDelta}_{21} \varDelta_{33} &
\overline{\varDelta}_{31} \varDelta_{33} \\
\overline{\varDelta}_{13} \varDelta_{12} &
\overline{\varDelta}_{23} \varDelta_{22} &
\overline{\varDelta}_{33} \varDelta_{22} &
\overline{\varDelta}_{23} \varDelta_{32} &
\overline{\varDelta}_{33} \varDelta_{32} \\
\overline{\varDelta}_{12} \varDelta_{11} &
\overline{\varDelta}_{22} \varDelta_{21} &
\overline{\varDelta}_{32} \varDelta_{21} &
\overline{\varDelta}_{22} \varDelta_{31} &
\overline{\varDelta}_{32} \varDelta_{31} \\
\overline{\varDelta}_{13} \varDelta_{11} &
\overline{\varDelta}_{23} \varDelta_{21} &
\overline{\varDelta}_{33} \varDelta_{21} &
\overline{\varDelta}_{23} \varDelta_{31} &
\overline{\varDelta}_{33} \varDelta_{31} \\
\end{bmatrix} ,
\end{equation}
\begin{equation}
B'_{diag}=
\begin{bmatrix}
\overline{\varDelta}_{11} \varDelta_{11} &
\overline{\varDelta}_{21} \varDelta_{21} &
\overline{\varDelta}_{31} \varDelta_{21} &
\overline{\varDelta}_{21} \varDelta_{31} &
\overline{\varDelta}_{31} \varDelta_{31} \\
\overline{\varDelta}_{12} \varDelta_{12} &
\overline{\varDelta}_{22} \varDelta_{22} &
\overline{\varDelta}_{32} \varDelta_{22} &
\overline{\varDelta}_{22} \varDelta_{32} &
\overline{\varDelta}_{32} \varDelta_{32} \\
\overline{\varDelta}_{13} \varDelta_{13} &
\overline{\varDelta}_{23} \varDelta_{23} &
\overline{\varDelta}_{33} \varDelta_{23} &
\overline{\varDelta}_{23} \varDelta_{33} &
\overline{\varDelta}_{33} \varDelta_{33} \\
\end{bmatrix}.
\end{equation}
The matrices occurring in (\ref{eq:culd}) have a form
\begin{equation}
C_{up}=
\begin{bmatrix}
\overline{\varGamma}_{11} \varGamma_{12} &
\overline{\varGamma}_{21} \varGamma_{22} &
\overline{\varGamma}_{31} \varGamma_{32} \\
\overline{\varGamma}_{11} \varGamma_{13} &
\overline{\varGamma}_{21} \varGamma_{23} &
\overline{\varGamma}_{31} \varGamma_{33} \\
\overline{\varGamma}_{12} \varGamma_{13} &
\overline{\varGamma}_{22} \varGamma_{23} &
\overline{\varGamma}_{32} \varGamma_{33} \\
\end{bmatrix} ,
\end{equation}
\begin{equation}
C_{low}=C_{up}(\varGamma \leftrightarrow \overline{\varGamma}),
\end{equation}
\begin{equation}
C_{diag}=
\begin{bmatrix}
\overline{\varGamma}_{11} \varGamma_{11} &
\overline{\varGamma}_{21} \varGamma_{21} &
\overline{\varGamma}_{31} \varGamma_{31} \\
\overline{\varGamma}_{12} \varGamma_{12} &
\overline{\varGamma}_{22} \varGamma_{22} &
\overline{\varGamma}_{32} \varGamma_{32} \\
\overline{\varGamma}_{13} \varGamma_{13} &
\overline{\varGamma}_{23} \varGamma_{23} &
\overline{\varGamma}_{33} \varGamma_{33} \\
\end{bmatrix}.
\end{equation}
Similarly we can write matrices from (\ref{eq:duld}) as
\begin{equation}
D_{up}=
\begin{bmatrix}
\overline{\varOmega}_{11} \varOmega_{12} &
\overline{\varOmega}_{21} \varOmega_{22} &
\overline{\varOmega}_{31} \varOmega_{32} \\
\overline{\varOmega}_{11} \varOmega_{13} &
\overline{\varOmega}_{21} \varOmega_{23} &
\overline{\varOmega}_{31} \varOmega_{33} \\
\overline{\varOmega}_{13} \varOmega_{13} &
\overline{\varOmega}_{23} \varOmega_{23} &
\overline{\varOmega}_{33} \varOmega_{33} \\
\end{bmatrix} ,
\end{equation}
\begin{equation}
D_{low}=
\begin{bmatrix}
\overline{\varOmega}_{12} \varOmega_{11} &
\overline{\varOmega}_{22} \varOmega_{21} &
\overline{\varOmega}_{32} \varOmega_{31} \\
\overline{\varOmega}_{13} \varOmega_{11} &
\overline{\varOmega}_{23} \varOmega_{21} &
\overline{\varOmega}_{33} \varOmega_{31} \\
\overline{\varOmega}_{13} \varOmega_{12} - \overline{\varOmega}_{12} \varOmega_{13} &
\overline{\varOmega}_{23} \varOmega_{22} - \overline{\varOmega}_{22} \varOmega_{23} &
\overline{\varOmega}_{33} \varOmega_{32} - \overline{\varOmega}_{32} \varOmega_{33} \\
\end{bmatrix},
\end{equation}
\begin{equation}
D_{diag}=
\begin{bmatrix}
\overline{\varOmega}_{11} \varOmega_{11} &
\overline{\varOmega}_{21} \varOmega_{21} &
\overline{\varOmega}_{31} \varOmega_{31} \\
\overline{\varOmega}_{12} \varOmega_{12} &
\overline{\varOmega}_{22} \varOmega_{22} &
\overline{\varOmega}_{32} \varOmega_{32} \\
(\overline{\varOmega}_{13} \varOmega_{12} + \overline{\varOmega}_{12} \varOmega_{13})/2 &
(\overline{\varOmega}_{23} \varOmega_{22} + \overline{\varOmega}_{22} \varOmega_{23})/2 &
(\overline{\varOmega}_{33} \varOmega_{32} + \overline{\varOmega}_{32} \varOmega_{33})/2 \\
\end{bmatrix}.
\end{equation}
and the matrices from (\ref{eq:dpuld}) look like
\begin{equation}
D'_{up}=
\begin{bmatrix}
\overline{\varOmega}_{11} \varOmega_{12} &
\overline{\varOmega}_{21} \varOmega_{22} &
\overline{\varOmega}_{31} \varOmega_{32} \\
\overline{\varOmega}_{11} \varOmega_{13} &
\overline{\varOmega}_{21} \varOmega_{23} &
\overline{\varOmega}_{31} \varOmega_{33} \\
\overline{\varOmega}_{12} \varOmega_{11} &
\overline{\varOmega}_{22} \varOmega_{21} &
\overline{\varOmega}_{32} \varOmega_{31} \\
\end{bmatrix} ,
\end{equation}
\begin{equation}
D'_{low}=D'_{up}(\varOmega \leftrightarrow \overline{\varOmega}),
\end{equation}
\begin{equation}
D'_{diag}=
\begin{bmatrix}
\overline{\varOmega}_{11} \varOmega_{11} &
\overline{\varOmega}_{21} \varOmega_{21} &
\overline{\varOmega}_{31} \varOmega_{31} \\
\overline{\varOmega}_{12} \varOmega_{12} &
\overline{\varOmega}_{22} \varOmega_{22} &
\overline{\varOmega}_{32} \varOmega_{32} \\
\overline{\varOmega}_{13} \varOmega_{12} &
\overline{\varOmega}_{23} \varOmega_{22} &
\overline{\varOmega}_{33} \varOmega_{32} \\
\overline{\varOmega}_{12} \varOmega_{13} &
\overline{\varOmega}_{22} \varOmega_{23} &
\overline{\varOmega}_{32} \varOmega_{33} \\
\overline{\varOmega}_{13} \varOmega_{13} &
\overline{\varOmega}_{23} \varOmega_{23} &
\overline{\varOmega}_{33} \varOmega_{33} \\
\end{bmatrix}.
\end{equation}

\mysection{Manifolds determined by set of $M$ equations in $N$
dimensional space}

\subsection{\label{sec:mz}Method of finding zeros
of the function $\vec{F}$ in~$N$ dimensions}

We shall be looking for solutions of a set equations
\begin{equation}
\label{eq:F0}
F_j(x_1,x_2,\ldots,x_N)=0 \ \ \ j=1,2,\ldots,M.
\end{equation}
Let $\vec{x}$ denote the  vector of values $x_i$, and
$\vec{F}$ the  vector of functions $F_j$. Let us expand
$F_j(\vec{x})$ in a Taylor series
\begin{equation}
\label{eq:taylor}
F_j( x_1+\delta x_1, x_1+\delta x_1,\ldots,x_N+\delta x_n)
=F_j(x_1, x_2, \ldots x_N)+\sum_{i=1}^{N}\px{F_j}{x_i}
\delta x_i + O(\delta x^2).
\end{equation}
The matrix of partial derivatives appearing in equation (\ref{eq:taylor})
is the rectangular Jacobian matrix ${\bf J}$.
Thus, in matrix notation equation (\ref{eq:taylor}) reads:
\begin{equation}
\vec{F}(\vec{x}+\delta \vec{x})=\vec{F}(\vec{x})+{\bf J}\delta \vec{x}
+ O(\delta \vec{x}^{~2}).
\label{eq:mtaylor}
\end{equation}
We are interested in zeros of $\vec{F}(\vec{x})$,
{\em i.e.} we are looking for such $\delta \vec{x}$
that $\vec{F}(\vec{x}+\delta \vec{x})=0$.
Neglecting terms of the order $ O(\delta \vec{x}^{~2})$,
we obtain a set of linear equations for the corrections
$\delta \vec{x}$
\begin{equation}
 {\mathbf J} \delta \vec{x}=- \vec{F}.
\label{eq:syseq}
\end{equation}

Equation (\ref{eq:syseq}) describes a set of $M$ linear equations
with $n$ values of solution $\delta \vec{x}$.
Each of these equations defines $(N-1)$ dimensional plane in $N$ dimensional space.
In order to solve (\ref{eq:syseq}) we should find
the intersection of these $N-1$-planes.

One should consider three cases:
\begin{enumerate}
\item If the number of linearly independent equations is equal to
the number of coordinates $M=N$ then the set of equations has only
one solution $\delta \vec{x}$;
\item If the number of linearly independent equations is lower than
the number of coordinates $M<N$
and the equations are not contradictory,
the set (\ref{eq:syseq}) has an infinite number of solutions which
form a $N-M$ dimensional plane. Then one selects the solution
from this $N-M$ plane which has the lowest norm.
\item In the case when $\delta \vec{x}$ doesn't exist, which means that
the equations are contradictory, we adopt a procedure which tries to
find some $\delta \vec{x}^{\;\prime}$
which decreases the test function (\ref{eq:mtaylor}).
If our set of equations is not contradictory then
the algorithm reduces itself to two other cases.
\begin{enumerate}
\item
Let $m$ be a number of linearly independent equations which are
not contradictory.
Then, from the set of equations (\ref{eq:syseq}) we can construct
$k$ sets of $m$ linearly independent equations.
Each of these sets determines $m$ entries of the $N$ dimensional
vector $\delta {\vec{x}}^{(i)\;\prime}$.
Here $i$ $(=1,\ldots,k)$ corresponds to the $i^{th}$ set of equations.
\item The remaining $(N-m)$ entries are set in a such way that
 $\delta \vec{x}^{(i)}{}'$ has the lowest norm.
\item Next we choose such $j$ that sum of the angles between a vector
 $\delta \vec{x}^{(j)}{}'$ and remaining vectors $\vec{x}^{(i)}{}'$
 is minimal. $\delta \vec{x}^{(j)}{}'$ has the nearest direction
to the average direction of other vectors $\vec{x}^{(i)}{}'$.
\end{enumerate}
\end{enumerate}

Instead of finding a zero of the $M$-dimensional function $\vec{F}$
we shall be looking for a global minimum of a function
\begin{equation}
f=\frac{1}{2}\vec{F}\cdot\vec{F}.
\label{eq:ff}
\end{equation}
Of course there can be some local minima of Eq.(\ref{eq:ff}) that are
not solutions of Eq.(\ref{eq:syseq}).

Our step $\delta \vec{x}$ is usually in the descent direction of $f$
\begin{equation}
\vec{\nabla} f \cdot \delta \vec{x}=(\vec{F} {\bf J})\cdot
(-{\bf J}^{-1}  \vec{F})=-\vec{F}\cdot\vec{F}<0
\end{equation}
which is true if only ${\bf J}^{-1}$ exists.

It is convenient to define
\begin{equation}
g(\lambda)\equiv f(\vec{x}_{old} + \lambda \vec{p})
\end{equation}
where $\vec{p}=\delta \vec{x}$.
We always first try the full step {\em i.e.} $\lambda=1$. If
the proposed step does not reduce $f$ we backtrack along the
same direction until we have an acceptable step
\begin{equation}
\vec{x}_{new}= \vec{x}_{old} + \lambda \vec{p}, \;\;\; 0<\lambda \le 1,
\end{equation}
{\em i.e.} we look for $\lambda$  which sufficiently reduces $g(\lambda)$.
This is done by approximating $g(\lambda)$ by polynomials in $\lambda$.
Initially we know $g(0)$ and $g'(0)$ and also $g(1)$ which is known
from the first trial ($\lambda=1$). Since
we can easily calculate the derivative of $g(\lambda)$
\begin{equation}
g'(\lambda)=\vec{\nabla} f \cdot \vec{p}.
\end{equation}
we can approximate $g(\lambda)$ by a quadratic polynomial
in $\lambda$:
\begin{equation}
g(\lambda)\simeq \left[g(1)-g(0)-g'(0) \right] \lambda^2 + g'(0) \lambda + g(0)
\label{eq:g}
\end{equation}
and look for the minimum of (\ref{eq:g}).
If this step also fails, we model $g(\lambda)$ as a cubic
polynomial in $\lambda$ and so on until the satisfactory value
of $\lambda$ is found.

It is obvious that because of the linearity of the algorithm
\begin{equation}
\vec{F}(\vec{x}+\delta \vec{x}) = O(\delta \vec{x}^2).
\end{equation}
But, if $f(\vec{x}+ \delta \vec{x})\le f(\vec{x})$,
then our procedure leads towards the solution of (\ref{eq:F0}),
provided we are not in a vicinity of a false (i.e. local) minimum of $f$.
In the latter case we have to change the initial conditions and
start be whole procedure again.

\subsection{Algorithm for finding the curves}

We shall describe a curve as a set of points placed along some
path where the distance $r$ between all neighboring points, should
be constant. By definition all points should be zeros of function
$\vec{F}$.

\begin{enumerate}
\item The input data of our algorithm are:
\begin{itemize}
\item[--] two points $\vec{y}_1$, $\vec{y}_2$,
which should be placed in the vicinity of the sought curve,
\item[--] distance $r$ between these adjacent points.
\end{itemize}
\item Making use of the algorithm from Section \ref{sec:mz}
we find a root $\vec{x}_1$, situated in the vicinity of $\vec{y}_1$.
\item We define the point $\vec{x}_1$ as a center of a
hyperspherical coordinate system. Next, we look
on the sphere with radius $r$ for the root $\vec{x}_2$
which has similar coordinates as point $\vec{y}_2$.
\item We shift the center of the coordinate system to the point $\vec{x}_2$
and look for a zero of $\vec{F}$ on the hypersphere of radius $r$
in vicinity of point $\vec{y}_k(k=3)$ extrapolated from previously
found roots $\vec{x}_i$ with $i=1,2,\ldots, k-1$.
\item Iterating the above procedure we construct
a curve of zeros of $\vec{F}$.
\end{enumerate}

Using this algorithm we can find not only curves, but also
$k$ dimensional hypersurfaces. This can be done by fixing
the values of $k-1$ coordinates.
Then for the each choice depending on which coordinates were fixed,
we can find curves from which we can in principle
reconstruct the hypersurface. However, in our case it turned out
that the zeros of $\vec{F}$ lie on one dimensional curves.

\end{document}